\documentclass{article}

\usepackage[a4paper, total={6in, 8in}]{geometry}
\usepackage[utf8]{inputenc}
\usepackage[T1]{fontenc} 
\usepackage{authblk}
\usepackage{url}
\usepackage{graphicx} 
\usepackage{bm}
\usepackage{amssymb}
\usepackage{amsmath}
\usepackage{braket}
\usepackage{subcaption}
\usepackage{sidecap}
\usepackage{tikz}
\usetikzlibrary{calc}
\usetikzlibrary{arrows.meta, decorations.pathreplacing}

\title{Slice-Wise Initial State Optimization to Improve Cost and Accuracy of the VQE on Lattice Models}
\author[1]{Cedric Gaberle}
\author[1]{Manpreet S. Jattana}
\affil[1]{Modular Supercomputing and Quantum Computing, Institute of Computer Science, Goethe University Frankfurt, D-60325 Frankfurt, Germany}
\date{}

\begin{document}

\maketitle

\begin{abstract}
    
    We propose an optimization method for the Variational Quantum Eigensolver (VQE) that combines adaptive and physics-inspired ansatz design. Instead of optimizing multiple layers simultaneously, the ansatz is built incrementally from its operator subsets, enabling subspace optimization that provides better initialization for subsequent steps. This quasi-dynamical approach preserves expressivity and hardware efficiency while avoiding the overhead of operator selection associated with adaptive methods. Benchmarks on one- and two-dimensional Heisenberg and Hubbard models with up to 20 qubits show improved fidelities, reduced function evaluations, or both, compared to fixed-layer VQE. The method is simple, cost-effective, and particularly well-suited for current noisy intermediate-scale quantum (NISQ) devices.
\end{abstract}

\section{Introduction}\label{sec:introduction}

As quantum hardware continues to advance, the development and improvement of practical quantum algorithms for near-term devices have gained increasing attention. While the current devices are still in the noisy intermediate-scale quantum (NISQ) era, the Variational Quantum Eigensolver (VQE), which overcomes the downsides of these devices in a classical-assisted optimization scheme, is one of the most promising algorithms to leverage quantum computing even in these still early stages of hardware development \cite{VQE_Review-of-methods-and-best-practices, VQE-on-photonic-quantum-processor, willsch2022}. It is one of the leading candidates to demonstrate quantum advantage through integration into existing High-Performance Computing systems  \cite{quantum_integration_survey, ompss-2, qaim}.

VQE is a hybrid quantum-classical algorithm designed to approximate the ground state, i.e., the lowest eigenstate, of a given Hamiltonian. A parameterized quantum circuit, called the ansatz, prepares a trial wavefunction that is optimized in a classical optimization loop by iteratively minimizing the expectation value of the energy. The capability of finding or closely approximating the true ground state of a system is highly dependent on the structure of the energy landscape and, consequently, the expressibility of the ansatz. Endeavors have been taken to balance expressivity, trainability, and hardware efficiency by optimizing the ansatz in use. Some developed physics-inspired ansätze, incorporating physically interesting phenomena and symmetries of the problem Hamiltonian into the ansatz to increase its expressivity \cite{VQE-on-photonic-quantum-processor, UCC, UCCSD, HVA2}. Such ansätze can increase significantly in size and computational effort needed to optimize \cite{VQE_Review-of-methods-and-best-practices}. Others develop the ansatz in a dynamic fashion during the VQE optimization, like the Adapt-VQE and its variants \cite{adaptVQE, qubit-adaptVQE, tetris-adaptVQE, qeb-adapt-vqe}. The adaptive methods construct the ansatz from a predefined set of meaningful operators by an operator-per-operator increase after each sequentially performed VQE optimization, ensuring an optimal depth of the resulting circuit with respect to a given accuracy of the results. The downside of an added optimization scheme is the increase in total computational effort for the VQE algorithm. 

In this work, we explore a middle ground between the two ansatz design techniques by combining the iterative increase in ansatz operators in use with the physically motivated ansätze developed in prior works \cite{HVA, QuasiDynamics}. The structure of the final ansatz is given by a physics-inspired ansatz, such as the Hamiltonian Variational Ansatz, which is inspired by the Trotterized decomposition of the time-evolution operator. This ansatz captures the symmetries of the given Hamiltonian and offers high expressivity due to repeated layers of unitary operator sequences that mimic time-evolved correlations. The ansatz is broken into multiple pieces serving as the operator building blocks which are added sequentially in order given by the ansatz, circumventing the problem of operator decision. Parameters of each block are optimized in isolation, with the optimal values being fixed in the next iteration serving as an improved initial state. The pre-optimization of lower-dimensional sub-regions of the energy landscape allows the full ansatz to span a different region of the landscape than starting with the regular initial state, taken from the best classical approximation known. The operator pool creation by ansatz slicing is arbitrary. The single-layer ansatz may be sliced $i \in \{0, \dots, n-1\}$ times, where $n$ is the number of single-parameter operations present in one layer. While the resulting depth, determined by the number of layers taken into account, may be larger than the optimal depth found by the adaptive method, the decision of the next operator to span a different subspace is at no additional cost, compared to the gradient calculation for each operator in the pool at every optimization step during the Adapt-VQE approach. 

The rest of the paper is structured as follows. Section \ref{sec:background} introduces the Variational Quantum Eigensolver and the Heisenberg and Hubbard models in more detail. For both models, the used ansatz for the evaluation is presented and motivated. Section \ref{sec:method} explains the optimization method and categorizes it in the current state-of-the-art dynamic ansatz design landscape to improve the VQE. The evaluation is presented in Section \ref{sec:results}, split into the Heisenberg and Hubbard model results, before we conclude in the last section and provide an outlook on the current state and possible follow-up in the realm of VQE. 

\section{Theoretical Background}\label{sec:background}
\subsection{Variational Quantum Eigensolver}\label{sec:vqe}
The Variational Quantum Eigensolver (VQE) is a hybrid quantum-classical algorithm designed to find the ground state $\ket{\psi}$ of a system by systematically minimizing the expectation value of its Hamiltonian. VQE, originally introduced in the field of quantum chemistry \cite{VQE-on-photonic-quantum-processor}, is applicable to a wide variety of problems and is one of the most promising near-term quantum computing use cases due to its applicability to devices of the current NISQ era. It leverages the strengths of both classical and quantum computing in a dynamic, iterative approach and has been shown to be capable of utilizing both quantum computing paradigms, gate-based and adiabatic (annealing) \cite{triple_hybrid}.

VQE is composed of two parts: the classical and the quantum. The quantum part prepares a trial state, described by the vector $\ket{\Tilde{\psi}}$ in Hilbert space, given by a predefined parameterized ansatz $U(\bm{\theta}) = \prod U_l(\theta_l)$, and calculates the expectation value of the Hamiltonian $\hat{H}$ of interest:
\begin{equation}
    \braket{\hat{H}}_{\Tilde{\psi}} = \braket{\Tilde{\psi}|\hat{H}|\Tilde{\psi}},
\end{equation} where $\ket{\Tilde{\psi}}$ is normalized.
The classical part optimizes the parameters in the ansatz to minimize the energy associated with the trial state and feeds the optimized values back into the ansatz to reinitiate the energy expectation value calculation until convergence, i.e., $\ket{\Tilde{\psi}} \rightarrow \ket{\psi}$, or until a predefined criterion is met. Crucial for the algorithm's capability of finding the correct state corresponding to the lowest energy, i.e., the ground state, is the choice of the parameterized ansatz $U(\bm{\theta})$, the initial state $\ket{\phi}$ of the quantum computer, and the optimization strategy used. A problem-inspired, classically obtained good initial point is usually translated into a preparation circuit to initialize the quantum device for the subsequent optimization:
\begin{align}
\begin{split}
    &\ket{\phi} = U_{\text{Prep}}\ket{0},\\
    &\ket{\Tilde{\psi}} = U(\bm{\theta})\ket{\phi},
\end{split}
\end{align}
As an example, in the context of the quantum Heisenberg model of spins, the N\'{e}el state refers to a configuration that represents the classical antiferromagnetic order. This means, on a lattice of spin-$\frac{1}{2}$ particles, in the N\'{e}el state adjacent spins alternate between opposite spin orientations, reflecting the minimum classical energy configuration for an antiferromagnetic Heisenberg Hamiltonian \cite{QuasiDynamics}. The optimization strategy can be broadly classified into several categories, such as gradient-based, gradient-free (derivative-free), and Bayesian optimization, each with its own strengths and weaknesses. The best fit is highly dependent on the problem's structure, the quality of the quantum computation and measurement, and the scalability. The parameterized ansatz is arguably the most important ingredient of the VQE algorithm. The ansatz dictates which regions of the Hilbert space can be explored. A good ansatz should be expressive, i.e., spanning a large region of the Hilbert space where the ground state resides, easy to optimize, and hardware-efficient. Most state-of-the-art ansätze incorporate known physical constraints or symmetries of the problem to solve, promising a more concise structure \cite{HVA, VQE-short-review, UCC}.

The right choice of ansatz is the primary factor to tackle the problem of barren plateaus, a phenomenon in optimization where the gradient vanishes exponentially with the size of the landscape, making optimization impossible \cite{barren-plateau-problem, barren-plateaus-gradient-free-optimization, barren-plateaus-qnn}. Different techniques have been explored to find a good ansatz for a given problem iteratively, but a universal ansatz suitable for a vast variety of problems unaffected by the barren plateau is unlikely to be found. The second most prominent problem in VQE is the noise and gate error of real devices. Hardware-efficient ansätze try to minimize the impact of imperfect quantum computation by fitting the ansatz implementation to the capabilities of the hardware \cite{hardware-efficient-usefulness, hardware-efficient-vqe}. Like for any other quantum computation, quantum error mitigation techniques can be used \cite{error-mitigation, error_mitigation2, error-mitigation-philip}. Insufficient accuracy of the computation and/or the readout limits the theoretical possibility to find the correct ground state and associated energy using VQE. 
\subsection{Hubbard Model}\label{sec:hubbard}
The Hubbard model, a fundamental theoretical model in condensed matter physics and quantum many-body theory, describes the behavior of interacting electrons in a lattice. It approximates the transition between conducting and insulating systems, providing insights into physical phenomena as magnetism, insulator transition, and superconductivity. While the Lieb-Wu Bethe ansatz is able to solve the one-dimensional Hubbard model exactly \cite{Lieb-Wu-Bethe}, higher dimensions are not solved in general. 

The fundamental concept of the Hubbard model is the description of the forces acting on electrons occupying the same lattice site (on-site Coulomb repulsion) and electrons of different lattice sites (kinetic energy). While the latter pushes the electrons to tunnel to neighboring sites, the on-site interaction pushes them away from their neighbors, therefore imposing competing forces on each electron in the lattice. Mathematically, this interaction can be described by the Hamiltonian
\begin{equation}
    \hat{H} = -t \sum_{\langle i,j \rangle} ( c_{i\sigma}^\dagger c_{j\sigma} + \text{h.c.}) + U \sum_i n_{i\uparrow}n_{i\downarrow},
\end{equation}
where $t$ denotes the hopping amplitude, $U$ the on-site interaction strength, $c_{i\sigma}^\dagger$ and $c_{i\sigma}$ are fermionic creation and annihilation operators of mode $\sigma_i \in \{\uparrow, \downarrow\}$ at site $i$, respectively, and $n_{i\sigma} = c_{i\sigma}^\dagger c_{i\sigma}$ is the number operator. 

\subsubsection*{Qubit Mapping}

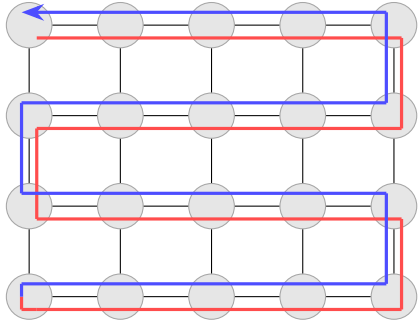
\begin{figure}
    \centering
    \begin{tikzpicture}[every node/.style={circle, draw=gray!70, fill=gray!20, minimum size=6mm},
                    >=Stealth]

\def\rows{3}
\def\cols{4}

\foreach \i in {0,...,\rows} {
    \foreach \j in {0,...,\cols} {
        \node (n\i-\j) at (\j*1.2,-\i*1.2) {};
    }
}

\foreach \i in {0,...,\rows} {
    \foreach \j in {0,...,\numexpr\cols-1} {
        \pgfmathtruncatemacro{\jnext}{\j + 1}
        \draw (n\i-\j) -- (n\i-\jnext);
    }
}

\foreach \i in {0,...,\numexpr\rows-1} {
    \foreach \j in {0,...,\cols} {
        \pgfmathtruncatemacro{\inext}{\i + 1}
        \draw (n\i-\j) -- (n\inext-\j);
    }
}

\newcommand{\belowcenter}[1]{($ (#1) + (0.1,-0.17) $)}  
\newcommand{\connectbelow}[1]{($ (#1) + (-0.1,-0.17) $)}  
\newcommand{\connectmiddle}[1]{($ (#1) + (-0.1,0) $)}  
\newcommand{\overcenter}[1]{($ (#1) + (-0.1,0.17) $)}  

\draw[very thick, red!70] \belowcenter{n0-0} -- \belowcenter{n0-4};
\draw[very thick, red!70] \belowcenter{n0-4} -- \belowcenter{n1-4};
\draw[very thick, red!70] \belowcenter{n1-4} -- \belowcenter{n1-0};
\draw[very thick, red!70] \belowcenter{n1-0} -- \belowcenter{n2-0};
\draw[very thick, red!70] \belowcenter{n2-0} -- \belowcenter{n2-4};
\draw[very thick, red!70] \belowcenter{n2-4} -- \belowcenter{n3-4};
\draw[very thick, red!70] \belowcenter{n3-4} -- \belowcenter{n3-0};
\draw[very thick, red!70] \belowcenter{n3-0} -- \connectbelow{n3-0};
\draw[very thick, red!70] \connectbelow{n3-0} -- \connectmiddle{n3-0};

\draw[very thick, blue!70] \connectmiddle{n3-0} -- \overcenter{n3-0};
\draw[very thick, blue!70] \overcenter{n3-0} -- \overcenter{n3-4};
\draw[very thick, blue!70] \overcenter{n0-4} -- \overcenter{n1-4};
\draw[very thick, blue!70] \overcenter{n1-4} -- \overcenter{n1-0};
\draw[very thick, blue!70] \overcenter{n1-0} -- \overcenter{n2-0};
\draw[very thick, blue!70] \overcenter{n2-0} -- \overcenter{n2-4};
\draw[very thick, blue!70] \overcenter{n2-4} -- \overcenter{n3-4};
\draw[->, very thick, blue!70] \overcenter{n0-4} -- \overcenter{n0-0};

\end{tikzpicture}
    \caption{Snake-like ordering of the Hubbard lattice's fermionic modes to qubits. Each site is represented by two qubits, one for each mode, for which the red line indicates spin-up qubit order and the blue line indicates spin-down qubit order.}
    \label{fig:snake-order}
\end{figure}

Since the Hubbard model describes fermions and the fermionic creation and annihilation operators obey anti-commutation relations, in order to simulate the Hubbard model using VQE, one needs to map the fermionic modes to qubits. We will focus on the Jordan-Wigner transformation \cite{jordan-wigner-transform}, but others like Bravyi-Kitaev and its variants could also be used \cite{bravyi-kitaev, bravyi-kitaev2}. It maps fermionic operators onto a combination of qubit operators written in terms of the Pauli matrices, effectively transforming the fermionic Hamiltonian into a spin Hamiltonian suitable for fermionic systems simulation on quantum computers. Each fermionic operator gets mapped to a qubit operator 
\begin{equation}\label{eq:spin_mapping}
\begin{aligned}
    c_{j, \sigma}^\dagger &\rightarrow \frac{1}{2}(X_{j_\sigma} - i Y_{j_\sigma}) \prod_{k=0}^{j_\sigma-1} Z_k, \\
    c_{j,\sigma} &\rightarrow \frac{1}{2} (X_{j_\sigma} + iY_{j_\sigma}) \prod_{k=0}^{j_\sigma-1} Z_k,
\end{aligned}
\end{equation}
where the strings of Pauli-Z operations enforce the fermionic anti-commutation relation and $j_\sigma$ is the index of the qubit representing mode $\sigma$ of site $j$. This implies that each site is represented by exactly two qubits, one for each spin. The mapping of qubit indices to lattice sites and modes is arbitrary. We decided upon a snake-like mapping, as illustrated in Fig. \ref{fig:snake-order}. According to Eq. \eqref{eq:spin_mapping}, the number operator is mapped onto qubit operators by the transformation
\begin{equation}
    n_{j, \sigma} = c_{j, \sigma}^\dagger c_{j, \sigma} \rightarrow \frac{1 - Z_{j_\sigma}}{2}.
\end{equation}
Using the above mappings, an arbitrary hopping and interaction term is mapped to qubit operators as such:
\begin{equation}
    \begin{aligned}
        &c_{i,\sigma}^\dagger c_{j,\sigma} + \text{h.c.} &&\rightarrow \frac{1}{2}(X_{i_\sigma} X_{j_\sigma} + Y_{i_\sigma} Y_{j_\sigma}) \prod_{k = i_\sigma+1}^{j_\sigma-1} Z_k, \\
        &n_{j,\uparrow}n_{j,\downarrow} &&\rightarrow \frac{1}{4}(1 - Z_{j\uparrow} - Z_{j\downarrow} + Z_{j\uparrow}Z_{j\downarrow}).
    \end{aligned}
\end{equation}

\subsubsection{Ansatz}\label{ch:hubbard_ansatz}

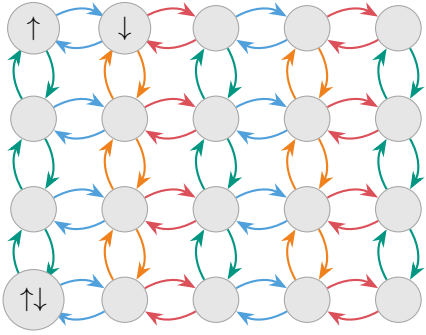
\begin{figure}
    \centering
    \begin{tikzpicture}[every node/.style={circle, draw=gray!70, fill=gray!20, minimum size=6mm},
                    >=Stealth]

\definecolor{myblue}{RGB}{80,160,220}   
\definecolor{myred}{RGB}{220,80,90}     
\definecolor{mygreen}{RGB}{0,150,130}
\definecolor{myorange}{RGB}{240,130,30}

\def\rows{3}
\def\cols{4}

\foreach \i in {0,...,\rows} {
    \foreach \j in {0,...,\cols} {
        \node (n\i-\j) at (\j*1.2,-\i*1.2) {};
    }
}

\foreach \i in {0,...,\rows} {
    \foreach \j in {0,...,\numexpr\cols-1} {
        \pgfmathtruncatemacro{\jnext}{\j + 1}
        \pgfmathtruncatemacro{\modulo}{mod(\j,2)}
        \ifnum\modulo=0
            \draw[->, thick, myblue] (n\i-\j) to[bend left=30] (n\i-\jnext);
            \draw[->, thick, myblue] (n\i-\jnext) to[bend left=30] (n\i-\j);
        \else
            \draw[->, thick, myred] (n\i-\j) to[bend left=30] (n\i-\jnext);
            \draw[->, thick, myred] (n\i-\jnext) to[bend left=30] (n\i-\j);
        \fi
    }
}

\foreach \i in {0,...,\numexpr\rows-1} {
    \foreach \j in {0,...,\cols} {
        \pgfmathtruncatemacro{\inext}{\i + 1}
        \pgfmathtruncatemacro{\modulo}{mod(\j,2)}
        \ifnum\modulo=0
            \draw[->, thick, mygreen] (n\i-\j) to[bend left=30] (n\inext-\j);
            \draw[->, thick, mygreen] (n\inext-\j) to[bend left=30] (n\i-\j);
        \else
            \draw[->, thick, myorange] (n\i-\j) to[bend left=30] (n\inext-\j);
            \draw[->, thick, myorange] (n\inext-\j) to[bend left=30] (n\i-\j);
        \fi
    }
}

\node (n0-0) at (0, 0) {$\uparrow$};
\node (n0-1) at (1.2, 0) {$\downarrow$};
\node (n3-0) at (0, -3.6) {$\uparrow\downarrow$};

\end{tikzpicture}
    \caption{Two-dimensional lattice of the Hubbard model. Arrows between the lattice sites indicate the hopping terms, and arrows on the sites indicate the spin occupation, which can either be non-occupied (empty node), single-occupied (upper left-hand side), or double-occupied (lower left-hand side). The kinetic term of the Hamiltonian can be divided into four parts, visualized by the different colors, containing only hopping terms between non-neighboring sites, i.e., horizontal at odd sites (blue), horizontal at even sites (red), vertical at odd sites (green), and vertical at even sites (orange). Hopping terms within each set mutually commute.}
    \label{fig:hubbard_lattice}
\end{figure}

Different ansätze are commonly used to solve the Hubbard model using variational algorithms, each with a focus on a different aspect. Some are more hardware-efficient, while others incorporate more physically motivated structures. We decided upon the Hamiltonian Variational Ansatz (HVA) \cite{HVA}, which can be attributed to the latter. 

The ansatz is constructed directly from the terms apparent in the system's Hamiltonian and is inspired by the adiabatic ground state evolution. Considering that the Hamiltonian can be grouped into $N\in \{2,\dots, 5\}$ groups, a maximum of 4 for the kinetic and one for the potential term (see Fig. \ref{fig:hubbard_lattice}), such that $ \hat{H} = \sum_{\alpha=1}^N H_\alpha$, the unitary operators of the HVA are composed of these grouped terms in a variational manner:
\begin{equation}\label{eq:hamiltonian_unitary}
    U(\bm{\theta}) = \prod_{\alpha=1}^N e^{i\theta_{\alpha}H_\alpha}.
\end{equation}
Moreover, similar to the adiabatic time evolution, the complete ansatz unitary incorporates $n$ steps of the unitary given in Eq. \eqref{eq:hamiltonian_unitary} to construct the HVA unitary
\begin{equation}
    U_{\text{HVA}}(\bm{\theta}) = \prod_{k=1}^n \prod_{\alpha = 0}^{N-1} e^{i\theta_{k,\alpha}H_\alpha} = \prod_{k=1}^n \prod_{\alpha = 1}^{N-1} e^{i\theta_{k,0}H_0}e^{i\theta_{k,\alpha}H_\alpha}.
\end{equation}
Comparing this to the unitary of the adiabatic time evolution under the influence of a Hamiltonian $\hat{H} = \hat{H}_0 + V$ with a non-interacting part $\hat{H}_0$ and interaction $V$
\begin{equation}
    U_{\text{ad}} = \prod_{k=1}^n e^{-i \frac{\tau}{n}\hat{H}_0} e^{-i \frac{\tau}{n}\frac{k}{n}V}
\end{equation}
where the time $\tau$ of the evolution is divided by a large number $n$ of Trotter steps, the similarity justifies the expectation of finding the full Hamiltonian's ground state $\ket{\psi_g}$ starting from the non-interacting ground state $\ket{\psi_0}$. Similar to the increased number of time steps in the adiabatic evolution, the HVA is expected to find more accurate results by increasing the number of layers $n$.
\subsection{Heisenberg Model}\label{sec:heisenberg}
The Heisenberg model is another foundational model in quantum many-body systems and condensed matter physics. The model describes the interaction of spins on a lattice. These spins are treated quantum mechanically, giving insights into the system's magnetic properties and quantum phase transitions.

At each site of the lattice, a spin $\sigma_i \in \{\pm 1\}$ representing a microscopic magnetic dipole resides, to which the magnetic moment is either up or down relative to a chosen axis, typically the z-axis. A common assumption is to include only nearest-neighbor interactions. 
In the Heisenberg model, spins are replaced by a Pauli spin operator ($\sigma^x, \sigma^y, \sigma^z$) and a choice of real-valued coupling constants in each direction:
\begin{equation}
    \hat{H} = \sum_{\langle i, j \rangle} ( J_x \sigma^x_i \sigma^x_j + J_y \sigma^y_i \sigma^y_j + J_z \sigma^z_i \sigma^z_j).
\end{equation}
Depending on the values of the coupling constants, different models are defined. We only consider the case $J_x = J_y = J_z = J$, termed the XXX model. The physics of the model depends on the sign of the coupling constant $J$, where the ground state is (anti-)ferromagnetic for a (positive) negative sign, respectively. The antiferromagnetic case is of greater interest to us, since it involves nontrivial entanglement and correlation, which makes it computationally harder to solve. Therefore, the final version of the Heisenberg Hamiltonian under study is given by:
\begin{equation}
    \hat{H} = J \sum_{\langle i,j \rangle} \bm{\sigma}_i \bm{\sigma}_j \: : \: \bm{\sigma} \in \{\sigma^x, \sigma^y, \sigma^z\}.
\end{equation}

\subsubsection{Ansatz}\label{ch:heisenberg_ansatz}
Despite the Bethe ansatz being able to find the exact wavefunction of the considered Heisenberg model in one dimension, therefore solving the ground state problem \cite{bethe-ansatz, karle2021bethe}, we use the VQE algorithm with an ansatz proposed in Ref. \cite{manpreet_heisenberg} in order to evaluate the proposed optimization method of this work. As mentioned previously, the choice of ansatz is crucial regarding whether the algorithm is capable of finding the correct ground state of the given Hamiltonian. It should have a large overlap with the ground state in question, spanning the relevant region of the considered Hilbert space, while still being computationally feasible to optimize classically.

The ansatz is given by
\begin{equation}\label{eq:heisenberg_ansatz}
    U(\bm{\theta}) = \prod_{\substack{l = N-1\\k=N}}^{\substack{k = l+1\\l=1}} U_{lk}(\theta_{lk}) U_{kl}(\theta_{kl}),
\end{equation}
where
\begin{equation}
    U_{kl} = \left\{ \begin{aligned}
        &e^{-i \theta_{kl} \sigma_k^y \sigma_l^x} &&\text{if $k = N$ or $l=N$,} \\
        &e^{-i \theta_{kl} \sigma_k^y \sigma_l^x \sigma_N^z} &&\text{otherwise.}
    \end{aligned}\right.
\end{equation}
This ansatz ensures that every combination of qubits has two unitary operations $U_{kl}$ acting upon them, where the $\sigma^x$ and $\sigma^y$ operators change between the higher and lower qubit index. The ansatz further introduces a $\sigma^z$ on the last qubit if not acted upon otherwise. This ensures that the parameterized rotation is always on the same qubit, allowing for easier mapping to a specific qubit on a real device. The ansatz produces low-depth circuits that can be implemented on current and future devices of the NISQ era, also taking into account that \textit{all-to-one} connectivity is easier to implement in contrast to \textit{all-to-all} connectivity for superconducting devices \cite{layerwise-learning-quantumNN}.

\begin{figure}
\centering
\begin{tikzpicture}[
    scale=1,
    site/.style={circle, fill=black, inner sep=1.2pt},
    bond/.style={line width=0.9pt}
]
    \def\R{1.4}   
    \def\out{1}   

    \coordinate (C) at (0,0);

    \foreach \i in {0,...,5}{
        \coordinate (H\i) at ({\R*cos(60*\i)},{\R*sin(60*\i)});
    }

    \foreach \i in {0,...,5}{
        \pgfmathtruncatemacro{\ip}{mod(\i+1,6)}
        \coordinate (M\i) at ($0.5*(H\i)+0.5*(H\ip)$);
        \coordinate (O\i) at ($ (M\i)!-\out!(C) $);
    }

    \foreach \i in {0,...,5}{
        \pgfmathtruncatemacro{\ip}{mod(\i+1,6)}
        \draw[bond,blue] (H\i)--(H\ip);
    }

    \foreach \i in {0,...,5}{
        \pgfmathtruncatemacro{\ip}{mod(\i+1,6)}
        \ifnum\i=1
        \else
            \ifnum\i=0
                \draw[bond,blue] (H\i)--(O\i) (O\i)--(H\ip);
            \else
                \ifnum\i=3
                    \draw[bond,blue] (H\i)--(O\i) (O\i)--(H\ip);
                \else
                    \ifnum\i=4
                        \draw[bond,green] (H\i)--(O\i) (O\i)--(H\ip);
                    \else
                     \ifnum\i=5
                      \draw[bond,blue] (H\i)--(O\i) (O\i)--(H\ip);
                     \else
                         \draw[bond,red] (H\i)--(O\i) (O\i)--(H\ip);
                        \fi¸
                    \fi
                \fi
            \fi
        \fi
    }

    \foreach \i in {0,...,5}{ \node[site] at (H\i) {}; }
    \foreach \i in {0,...,5}{
        \ifnum\i=1
        \else
            \node[site] at (O\i) {};
        \fi
    }

\end{tikzpicture}
\caption{Kagome lattice visualized for 9 sites (blue), 10 sites (+red), and 11 sites (+green). The qubits representing the sites are indexed from top to bottom, left to right.}
\label{fig:kagome_lattice}
\end{figure}
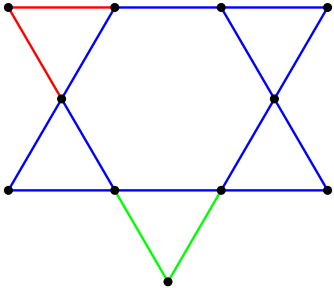

\section{Method}\label{sec:method}

As mentioned earlier, the ansatz selection is crucial to the performance and capability of the VQE algorithm in finding the ground state. To find the most optimal ansatz, dynamic approaches have been proposed, building the ansatz iteratively during the optimization loop. Such ansatz design techniques build on the idea of taking a predefined set of operators with physical meaning to the problem and decide only for the most promising during each optimization to add to the ansatz itself, therefore extending the ansatz on the fly. Such techniques include the Adapt-VQE and its variants \cite{adaptVQE, qubit-adaptVQE,qeb-adapt-vqe, tetris-adaptVQE}. Currently, the decision of the operator to append is based on its gradient at the current quantum state during the optimization process. Others have already pointed out that the gradient may not be the most decisive factor regarding the influence of the operator on the overall ansatz's capability of finding the ground state \cite{qeb-adapt-vqe, tetris-adaptVQE}. On real devices, calculating the gradient poses a critical challenge, e.g., due to the inherent shot noise caused by finite sampling. Moreover, the decision of the next best operator contributes to the overall runtime of the whole optimization process and requires a non-negligible number of additional queries to the quantum device. 

Other dynamical methods discard the operator selection and use known, well-performing ansätze and extend them by concatenating the same ansatz repeatedly. This is a Trotter-inspired technique which can be applied to either ansatz proposed in Sec. \ref{sec:background}. This technique was also used in Ref. \cite{QuasiDynamics} for the Heisenberg model and ansatz shown in Sec. \ref{ch:heisenberg_ansatz}. By increasing the number of layers, each layer representing the original unitary, the ansatz increases in its expressivity due to the increase in variational parameters. The resulting more flexible wave function explores added regions of the Hilbert space, therefore improving the ansatz's capability of approximating the ground state. For $n\rightarrow \infty$ and considering a noise-free simulation, the approximation of the ground state is postulated to become exact \cite{QuasiDynamics}. An infinite increase in ansatz depth and parameters is, nevertheless, of course, not feasible. 

Our method combines the two previously mentioned dynamical ansatz techniques by finding a middle ground between only single-parameter operators and whole ansatz unitaries of many parameters and operators being added after each optimization step. As the ansatz is not genuinely being constructed dynamically, we refer to the method as quasi-dynamical improved VQE. It leverages the strengths of adding only a smaller unitary to the ansatz at a time, similar to the adaptive dynamical methods, burrowing through a subspace of the parameter landscape during each iterative optimization and expand it only slowly to cover the Hilbert space sufficiently. Moreover, it preserves the advantages of the Trotter-inspired ansätze and their physical information, simplifying the pool creation and operator decision while ensuring the expressivity. If the problem at hand allows for a Trotter-inspired technique and a suitable good ansatz is already known, we use this ansatz as a layer in an iterative fashion but do not fix the overall structure beforehand. Assume, the single layer can be expressed by the unitary
\begin{equation}
    U_s(\bm{\theta}_s) = e^{-i \bm{\theta}_s P} = \prod_{k=1}^{|\bm{\theta}_s|} e^{-i \theta_k P_k},
\end{equation}
where $\bm{\theta}_s$ denotes the set of all parameters present in a single layer unitary and $P$ is the generator, e.g., a Pauli string. The $n$-layer ansatz is written as a product of single-layer unitaries
\begin{equation}
    U_n(\bm{\theta}_n) = \prod_{k=1}^n U_{s,k}(\bm{\theta}_{s,k}).
\end{equation}
Furthermore, $\bm{\theta}_{s,i} \neq \bm{\theta}_{s,j}$ for $i\neq j$ and $\bm{\theta}_n = \bigcup_k \bm{\theta}_{s,k}$. During the optimization, our technique appends a subset of the ansatz's operators as a single parameterized unitary at each step $t$, which can range in size from a single parameterized operator to a whole layer:
\begin{equation}
\begin{alignedat}{3}
    &U_{t=1}(\bm{\theta}_1) &&= \prod_{\theta_k \in \bm{\theta_1}} e^{-i \theta_k P_k}\ &&:\ \bm{\theta}_1 \subseteq \bm{\theta}_s, \\
    &U_{t{\scriptscriptstyle >}1}(\bm{\theta}_t) &&= \prod_{k=1}^{t-1} U_{t=k}(\bm{\vartheta}_k) e^{-i \bm{\theta}_t P_t}\ &&:\ \bm{\theta}_t \subseteq (\bm{\theta}_s \setminus \bigcup_{i{\scriptscriptstyle >}t} \bm{\theta}_i),
\end{alignedat}
\end{equation}
where $\bm{\vartheta}_k = (\vartheta_{k,1}, \vartheta_{k,2}, \dots)\ :\ \vartheta_{k,i} \in \mathbb{R}$ are the previously optimized and now fixed parameters of all steps until $t-1$. 
This allows for a subspace optimization, effectively altering the initial state, beneficial for the subsequent optimization. The ansatz can be sliced arbitrarily, resulting in a fixed operator pool from which to append. The decision of the unitary to add comes without the overhead of the gradient calculation, as in the Adapt-VQE, due to the predefined order given by the Trotter-inspired (multi-layer) ansatz. The unitaries in the operator pool are appended in order as they appear in the non-sliced ansatz. It follows that in the end $U_{t=t} = U_n$. The last step involves optimizing the full ansatz unitary, opening all parameters present. The difference lies in the initialization of those, since in our method all parameters have been pre-optimized on a subspace easier to navigate, therefore they already carry more information beneficial for the optimization of the full, possibly more rugged, parameter landscape. 

Despite not being as flexible as the adaptive methods, our method is more flexible than the Trotter-inspired technique alone, while circumventing the problem of operator decision and insufficient operator expressiveness. This reduces the associated overhead and makes use of existing physically meaningful ansätze and combines them with an improved optimization technique. As a quantum computer is a resource, like any other computational resource, one should utilize it optimally, i.e., reduce the required runtime to obtain sufficiently good results. In the next section, we show that the proposed optimization method improves either the accuracy of the result or the necessary queries to the device, or even both, marking it an easy adjustment to make to reduce complexity and increase usability of quantum devices.

\section{Results}\label{sec:results}

To evaluate the proposed quasi-dynamical optimization method, we conducted several experiments applying it to problems of the Heisenberg and Hubbard models of sizes up to 20 qubits. All experiments were quantum computing simulations utilizing the Qiskit Python library \cite{qiskit2024}. The minimization technique used in the parameter optimization step of the VQE is the gradient-based Broyden-Fletcher-Goldfarb-Shanno (BFGS) algorithm \cite{bfgs1, bfgs2, bfgs3, bfgs4} implemented by SciPy \cite{SciPy-NMeth}. The gradient tolerance is set to $10^{-5}$ for all problems and sizes investigated. In the following, we compare the ground state approximations found by the standard VQE and our quasi-dynamical improved VQE in terms of fidelity and function evaluations of the optimizer. The latter is no limiting factor for quantum computing simulation, but using a real device, every function evaluation directly translates into a query to the quantum computer.

\subsection{Hubbard Model}\label{sec:results_hubbard}

We consider different sizes and dimensions of the Hubbard model to compare the proposed method to the standard layer-wise HVA. For the one-dimensional case, the lattice (or chain) is of size $1\times 6$, $1\times 8$, or $1\times 10$ lattice sites. For the two-dimensional case, we investigated the lattice sizes $2\times 3$ and $3\times 3$. All considered sizes are not being solved to an accuracy of \textgreater99\% fidelity using only a single layer in the HVA, leaving room for improvement by an increase in layer combinations. The initial state is the ground state of the non-interacting case, defined by its Slater determinant and prepared by Givens rotations \cite{quantum-algos-for-many-body-physics, solving-fermi-hubbard}.

\begin{figure}
 \center
    \includegraphics[width=.5\linewidth]{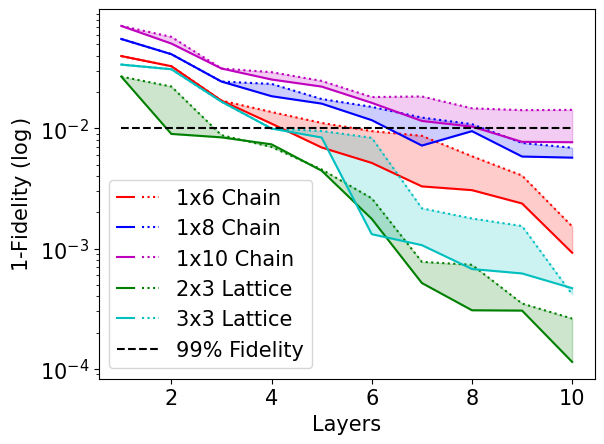}
    \caption{Relation of VQE fidelity with true ground state to number of layers/steps in the multi-layer ansatz of regular (dashed lines) and method-improved VQE (solid lines), respectively, for different lattice sizes. The larger the deviation between the dashed and solid lines, the greater the improvement of our method for the same number of layers, respectively, circuit depth. The 99\% fidelity rate is depicted by the black dashed line. Note that the y-axis depicts $1-$Fidelity, therefore, a lower value is favorable.}
    \label{fig:hubbard-fidelity}
\end{figure}

Figure \ref{fig:hubbard-fidelity} represents the fidelities for the different sizes of lattices and methods used. One can immediately see that an increase in the lattice size increases the complexity of the model for either dimension, as expected. Clearly noticeable is the deviation in fidelity improvement for an increase in ansatz layers with a favorable progression using our proposed method in all cases considered. Moreover, the number of layers necessary to reach \textgreater99\% fidelity is less, except for the $3\times 3$ lattice size, which reaches there at four layers for both. This implies that by using our method, shallower ansätze are capable of approximating the ground state more accurately. 

\begin{figure}
 \includegraphics[width=\linewidth]{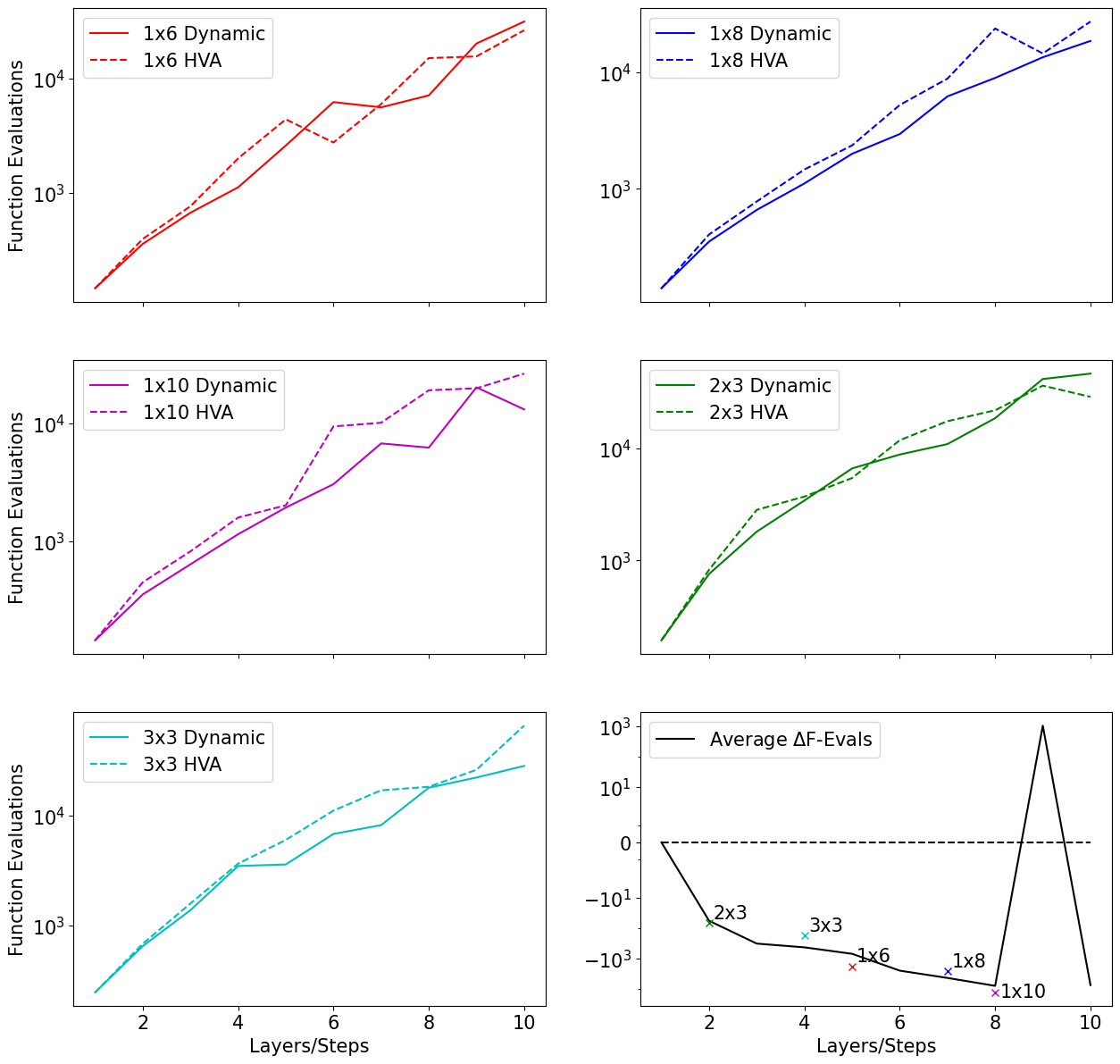}  
    \caption{Comparison of the number of function evaluations performed for different numbers of layers for all Hubbard model lattice sizes considered. The bottom right corner shows the average delta of function evaluations for our method compared to HVA, as well as the difference for each lattice size for the layer number necessary to achieve \textgreater99\% fidelity using our method.}
    \label{fig:hubbard_function_evals}
\end{figure}

Even considering that the batch-wise optimization of the different layer ansatz's parameters poses an additional optimization step in the overall VQE, we see in Fig. \ref{fig:hubbard_function_evals} that the number of function evaluations is, for most numbers of layers and problem sizes, less than using the full HVA directly. Only the nine-layer case experiences a surge in function evaluations for our method compared to standard VQE. The reason for the break in the trajectory remains unclear. One can also see that our enhanced method reaches \textgreater99\% fidelity for all sizes considered, using fewer function evaluations overall, i.e., the batch-wise optimization and full optimization at the end combined. This significantly reduces the queries to the quantum device, subsequently reducing the computational effort. This becomes even more significant when working with real devices rather than simulations, as inherent shot noise requires multiple repeated energy calculations to obtain a reasonably accurate approximation.

\subsection{Heisenberg Model}\label{sec:results_heisenberg}

For the Heisenberg model, we are considering one-dimensional chains of sizes $8-16$ sites and Kagome lattices of nine, ten, and eleven sites as depicted in Fig. \ref{fig:kagome_lattice}. The number of qubits necessary is the equivalent of the lattice sites of the problem. 

\begin{figure}
    \centering
    \includegraphics[width=.8\linewidth]{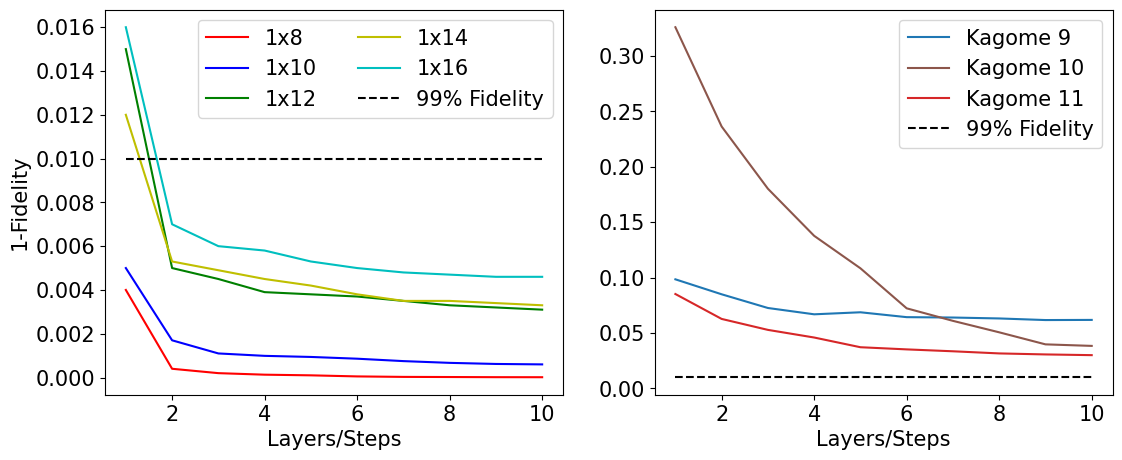}
    \caption{Increasing the layers of the Ansatz for the Heisenberg model Hamiltonians of sizes $1\times \{8, 10, 12, 14, 16\}$ and Kagome lattices of 9, 10, and 11 qubits. The dashed line indicates the 99\% fidelity threshold.}
    \label{fig:heisenberg_dynamical_all}
\end{figure}

First, investigating the behavior of repeated optimizations of the same ansatz successively, effectively altering the initial state of subsequent runs, Fig. \ref{fig:heisenberg_dynamical_all} shows that appending the same ansatz to the previously optimized solution improves the fidelity of the results found in all cases. This demonstrates the improvement possibilities using deeper optimization schemes and serves as a baseline to further investigate the proposed optimization method on the given problem sizes. Nevertheless, chains of sizes $\leq 1\times 10$ sites, while still benefitting from an increase in layers, are already solved with a fidelity \textgreater99\% using only a single layer, i.e., the ansatz itself is expressive enough to find a sufficiently good approximation starting from the N\'eel state. The Kagome lattices are unable to surpass the 99\% fidelity threshold even with ten iterations of ansatz optimization repetitions.

\begin{figure}
 \centering
 \includegraphics[width=.7\linewidth]{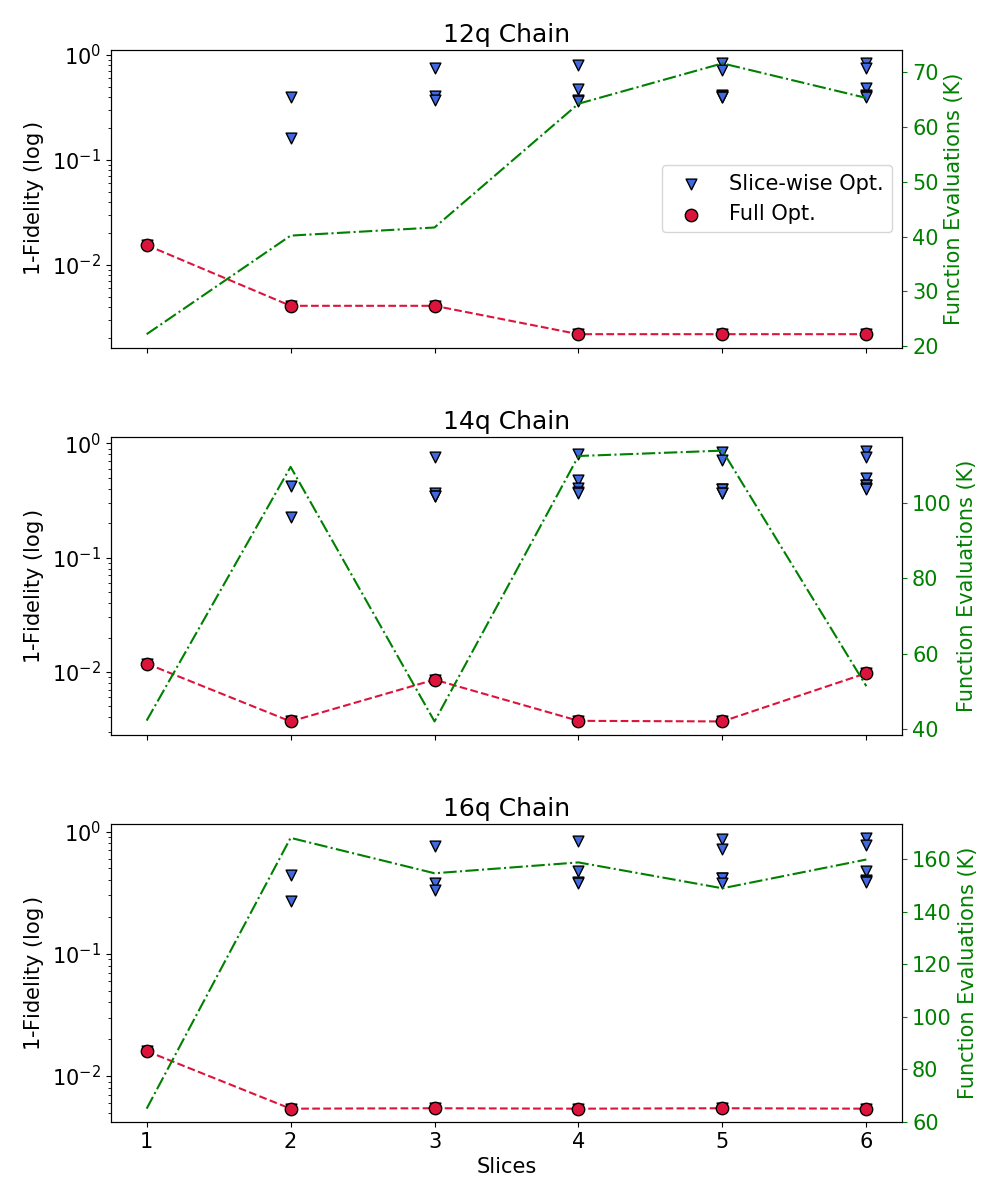}
 \caption{Different number of slices for the method-improved VQE on the Heisenberg chains of 12, 14, and 16 sites for a single-layer ansatz. The first entry (1 Slice) corresponds to the normal VQE. Blue triangles illustrate the fidelity of each slice added, the red dots the final optimization's fidelity.}
 \label{fig:heisenberg_heuristic}
\end{figure}

Investigating the 12, 14, and 16 sites Heisenberg chains, figure \ref{fig:heisenberg_heuristic} depicts the difference in fidelity and function evaluations for the normal VQE and the method-improved version for different numbers of slices of a single layer of the ansatz. The method-improved VQE is capable of improving the fidelities of all problem sizes, regardless. The fidelity is $>99\%$ for all numbers of slices for either problem size, but it is important to note that the normal VQE was already close to surpassing this threshold with only a single-layer ansatz, as can be seen in Fig. \ref{fig:heisenberg_dynamical_all}. The function evaluations, on the other hand, are always higher than the normal VQE counterpart. This indicates that the fidelity improvement comes with an increase in function evaluations by a factor of roughly $2.5$ for the best fidelity rate, respectively.

\begin{figure}
    \centering
    \includegraphics[width=\linewidth]{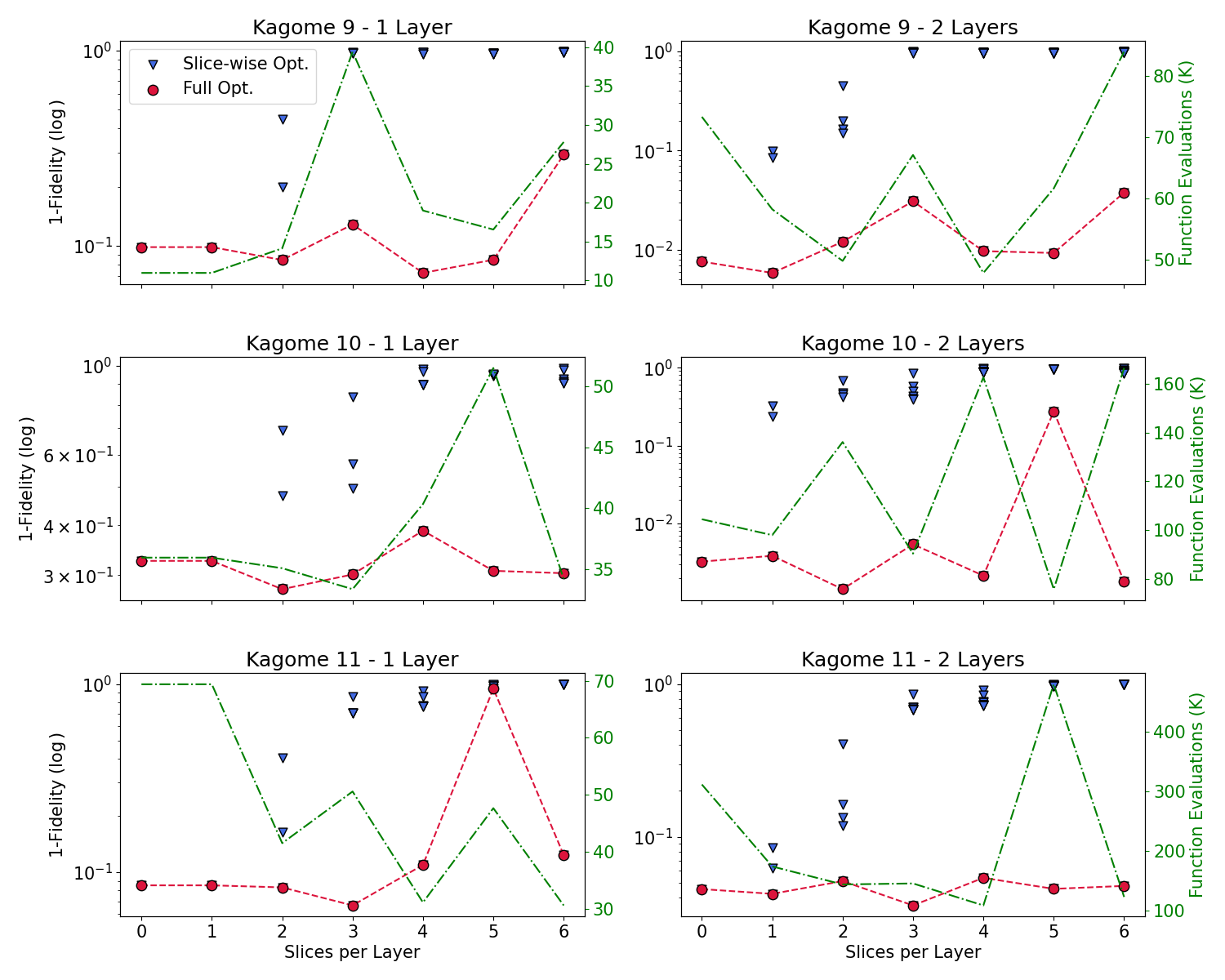}
    \caption{Method-improved VQE on the Heisenberg Kagome lattices for a single- and two-layer ansatz and different numbers of slices per layer. Blue triangles indicate the fidelity during slice-wise optimization, and red dots indicate the full optimization starting from the method-improved parameters. Zero slices means taking the ansatz as it is, while $x$ slices means separating each layer into $x$ parts, building, and optimizing the ansatz successively. Essentially, for the single-layer ansatz, 0 and 1 slices per layer mean the same. The green line indicates the total number of function evaluations performed until convergence.}
    \label{fig:kagome_heuristic1-2}
\end{figure}

Figure \ref{fig:kagome_heuristic1-2} compares the single-layer ansatz, the two-layer ansatz, and the method-improved versions of both for the Kagome lattices. It is apparent that the single-layer ansatz does not solve the problem satisfactorily for either size, while the two-layer ansatz can be improved to find solutions $>99$\% fidelity for the nine and ten-site kagome lattices. Still, the method-improved VQE is capable of improving the fidelities for all cases, but not for all numbers of slices per layer. The right decision of slices plays a crucial role in whether an improvement can be expected. The results suggest that respecting the structure of the ansatz when partitioning is beneficial. For example, the partitioning into two slices per ansatz for the ten-site kagome is performing the best. Since the ansatz is symmetrical, partitioning into two slices per layer equals a separation into the unitaries shown in \ref{ch:heisenberg_ansatz}. This is a reasonable, but not generally the best, partition for all experiments. Another reasonable separation for the two-layer cases is to slice into the single-layer ansätze during the method's subspace optimization. This yields the best result for the two-layer nine-site Kagome. The optimal number of slices cannot be specified universally and depends strongly on the problem. An indefinite increase in partitions does not necessarily equal an improvement.

In Fig. \ref{fig:kagome_heuristic1-2}, one can also see the behavior of the fidelity of the state found by each added part of the ansatz. The blue dots indicate that for each slice, extending the ansatz improves the fidelity of the optimized state, ultimately serving as an improved initial state for the full optimization, illustrated by the red dots. This was to be expected, given the insights from the results shown in Fig. \ref{fig:heisenberg_dynamical_all}. One can also see that the smaller the slices added in each step, the smaller the gain of initial state fidelity for subsequent optimizations.

Furthermore, Figure \ref{fig:kagome_heuristic1-2} shows the number of function evaluations performed in total until convergence. For the single layer, not only does the method-improved two-slices-per-layer version outperform the normal VQE, but the number of function evaluations to converge to the better result is also less. We can also see that the method struggles when the slice optimization is unable to improve the next initial state, i.e., the blue dots are close together. Then either the fidelity or the number of evaluations deteriorates. This is especially evident in the two-layer approach. The larger fidelity improvement relative to the added function evaluations in the two-slice-per-layer case supports the claim that the method outperforms standard VQE. Ultimately, the evaluation of improvement is to be determined by the usage and the capabilities of the device, i.e., to what extent an increase in fidelity justifies an increase in function evaluations. 

\section{Conclusion}\label{sec:conclusion}

In this work, we demonstrated how iteratively increasing a predefined ansatz by optimizing sub-ansätze in an adaptive fashion can beneficially influence the performance of the VQE on problems of the 1D and 2D Heisenberg and Hubbard models. The proposed improved heuristic fits right between fully adaptive methods and problem-inspired ansatz design techniques. By utilizing information of the problem in the ansatz decision, we are able to exploit symmetries and reduce the covered Hilbert space to a smaller interesting subspace. Additionally, adding the iterative approach of the sub-optimization routine, we can navigate the covered Hilbert space more efficiently, i.e., we can improve the overall results found or converge to an approximate solution faster. Evaluations of the cost function require calculating the energy expectation value using the quantum computer, which, due to hardware constraints and computational cost purposes, should be kept as few as possible while ensuring convergence. 

This approach bypasses the computational overhead associated with the operator selection of other adaptive methods to build the ansatz during the VQE optimization. Moreover, the ansatz is extended by a fixed number of operators and parameters to optimize in each step, with only a single full optimization at the end, additionally reducing the effort during the parameter optimization as a whole. While the implementation of the proposed improvement comes at minimal cost, the necessity of a good initial ansatz selection remains crucial. We also demonstrated that the incorporated symmetries of the problem in the ansatz should be considered when slicing into sub-ansätze. 

The improved VQE opens new avenues to circumvent current limitations of the algorithm, especially when using NISQ devices. Reducing the strain on these devices enables more efficient utilization compared to deeper full ansätze or adaptive methods, which require repeated measurements. Further investigation is needed into how (shot) noise influences the overall performance and the possibility of finding the ground state using the improved subspace optimized initialization VQE.

\bibliographystyle{ieeetr}
\bibliography{bibliography}

\begin{thebibliography}{10}

\bibitem{VQE_Review-of-methods-and-best-practices}
J.~Tilly, H.~Chen, S.~Cao, D.~Picozzi, K.~Setia, Y.~Li, E.~Grant, L.~Wossnig,
  I.~Rungger, G.~H. Booth, and J.~Tennyson, ``The variational quantum
  eigensolver: A review of methods and best practices,'' {\em Physics Reports},
  vol.~986, pp.~1--128, 2022.
\newblock The Variational Quantum Eigensolver: a review of methods and best
  practices.

\bibitem{VQE-on-photonic-quantum-processor}
A.~Peruzzo, J.~McClean, P.~Shadbolt, M.-H. Yung, X.-Q. Zhou, P.~J. Love,
  A.~Aspuru-Guzik, and J.~L. O'Brien, ``A variational eigenvalue solver on a
  photonic quantum processor,'' {\em Nature Communications}, vol.~5, p.~4213,
  July 2014.

\bibitem{willsch2022}
D.~Willsch, M.~Jattana, M.~Willsch, S.~Schulz, F.~Jin, H.~D. Raedt, and
  K.~Michielsen, ``Hybrid quantum classical simulations,'' 2022.

\bibitem{quantum_integration_survey}
P.~Döbler and M.~S. Jattana, ``A survey on integrating quantum computers into
  high performance computing systems,'' 2025.

\bibitem{ompss-2}
P.~Döbler, D.~Álvarez, L.~J. Menger, T.~Lippert, V.~Beltran, and M.~S.
  Jattana, ``Extending the ompss-2 programming model for hybrid
  quantum-classical programming,'' 2025.

\bibitem{qaim}
Z.~Zhu, C.~Gaberle, S.~M. Neuwirth, T.~Lippert, and M.~S. Jattana, ``Q-aim: A
  unified portable workflow for seamless integration of quantum resources,''
  2025.

\bibitem{UCC}
S.~Guo, J.~Sun, H.~Qian, M.~Gong, Y.~Zhang, F.~Chen, Y.~Ye, Y.~Wu, S.~Cao,
  K.~Liu, C.~Zha, C.~Ying, Q.~Zhu, H.~Huang, Y.~Zhao, S.~Li, S.~Wang, J.~Yu,
  D.~Fan, D.~Wu, H.~Su, H.~Deng, H.~Rong, Y.~Li, K.~Zhang, T.~Chung, F.~Liang,
  J.~Lin, Y.~Xu, L.~Sun, C.~Guo, N.~Li, Y.~Huo, C.~Peng, C.~Lu, X.~Yuan,
  X.~Zhu, and J.~Pan, ``Experimental quantum computational chemistry with
  optimized unitary coupled cluster ansatz,'' {\em Nature Physics}, vol.~20,
  pp.~1240--1246, June 2024.

\bibitem{UCCSD}
C.~Hempel, C.~Maier, J.~Romero, J.~McClean, T.~Monz, H.~Shen, P.~Jurcevic,
  B.~P. Lanyon, P.~Love, R.~Babbush, A.~Aspuru-Guzik, R.~Blatt, and C.~F. Roos,
  ``Quantum chemistry calculations on a trapped-ion quantum simulator,'' {\em
  Phys. Rev. X}, vol.~8, p.~031022, Jul 2018.

\bibitem{HVA2}
D.~Wecker, M.~B. Hastings, and M.~Troyer, ``Progress towards practical quantum
  variational algorithms,'' {\em Phys. Rev. A}, vol.~92, p.~042303, Oct 2015.

\bibitem{adaptVQE}
H.~R. Grimsley, S.~E. Economou, E.~Barnes, and N.~J. Mayhall, ``An adaptive
  variational algorithm for exact molecular simulations on a quantum
  computer,'' {\em Nature Communications}, vol.~10, p.~3007, July 2019.

\bibitem{qubit-adaptVQE}
H.~L. Tang, V.~Shkolnikov, G.~S. Barron, H.~R. Grimsley, N.~J. Mayhall,
  E.~Barnes, and S.~E. Economou, ``Qubit-adapt-vqe: An adaptive algorithm for
  constructing hardware-efficient ans\"atze on a quantum processor,'' {\em PRX
  Quantum}, vol.~2, p.~020310, Apr 2021.

\bibitem{tetris-adaptVQE}
P.~G. Anastasiou, Y.~Chen, N.~J. Mayhall, E.~Barnes, and S.~E. Economou,
  ``Tetris-adapt-vqe: An adaptive algorithm that yields shallower, denser
  circuit ans\"atze,'' {\em Phys. Rev. Res.}, vol.~6, p.~013254, Mar 2024.

\bibitem{qeb-adapt-vqe}
Y.~S. Yordanov, V.~Armaos, C.~H.~W. Barnes, and D.~R.~M. Arvidsson‑Shukur,
  ``Qubit‑excitation‑based adaptive variational quantum eigensolver,'' {\em
  Communications Physics}, vol.~4, p.~228, Oct. 2021.

\bibitem{HVA}
R.~Wiersema, C.~Zhou, Y.~de~Sereville, J.~F. Carrasquilla, Y.~B. Kim, and
  H.~Yuen, ``Exploring entanglement and optimization within the hamiltonian
  variational ansatz,'' {\em PRX Quantum}, vol.~1, p.~020319, Dec 2020.

\bibitem{QuasiDynamics}
M.~S. Jattana, F.~Jin, H.~De~Raedt, and K.~Michielsen, ``Improved variational
  quantum eigensolver via quasidynamical evolution,'' {\em Phys. Rev. Appl.},
  vol.~19, p.~024047, Feb 2023.

\bibitem{triple_hybrid}
M.~S. Jattana, ``Quantum annealer accelerates the variational quantum
  eigensolver in a triple-hybrid algorithm,'' {\em Physica Scripta}, vol.~99,
  p.~095117, aug 2024.

\bibitem{VQE-short-review}
D.~A. Fedorov, B.~Peng, N.~Govind, and Y.~Alexeev, ``{VQE} method: a short
  survey and recent developments,'' {\em Materials Theory}, vol.~6, no.~1,
  p.~2, 2022.

\bibitem{barren-plateau-problem}
M.~Larocca, S.~Thanasilp, S.~Wang, K.~Sharma, J.~Biamonte, P.~J. Coles,
  L.~Cincio, J.~R. McClean, Z.~Holmes, and M.~Cerezo, ``Barren plateaus in
  variational quantum computing,'' {\em Nature Reviews Physics}, vol.~7,
  pp.~174--189, 2025.

\bibitem{barren-plateaus-gradient-free-optimization}
A.~Arrasmith, M.~Cerezo, P.~Czarnik, L.~Cincio, and P.~J. Coles, ``Effect of
  barren plateaus on gradient-free optimization,'' {\em Quantum}, vol.~5,
  p.~558, 2021.

\bibitem{barren-plateaus-qnn}
J.~R. McClean, S.~Boixo, V.~N. Smelyanskiy, R.~Babbush, and H.~Neven, ``Barren
  plateaus in quantum neural network training landscapes,'' {\em Nature
  Communications}, vol.~9, no.~1, p.~4812, 2018.

\bibitem{hardware-efficient-usefulness}
L.~Leone, S.~F. Oliviero, L.~Cincio, and M.~Cerezo, ``On the practical
  usefulness of the {H}ardware {E}fficient {A}nsatz,'' {\em {Quantum}}, vol.~8,
  p.~1395, July 2024.

\bibitem{hardware-efficient-vqe}
A.~Kandala, A.~Mezzacapo, K.~Temme, M.~Takita, M.~Brink, J.~M. Chow, and J.~M.
  Gambetta, ``Hardware-efficient variational quantum eigensolver for small
  molecules and quantum magnets,'' {\em Nature}, vol.~549, pp.~242--246, Sept.
  2017.

\bibitem{error-mitigation}
S.~Endo, Z.~Cai, S.~C. Benjamin, and X.~Yuan, ``Hybrid quantum-classical
  algorithms and quantum error mitigation,'' {\em Journal of the Physical
  Society of Japan}, vol.~90, no.~3, p.~032001, 2021.

\bibitem{error_mitigation2}
M.~S. Jattana, F.~Jin, H.~De~Raedt, and K.~Michielsen, ``General error
  mitigation for quantum circuits,'' {\em Quantum Information Processing},
  vol.~19, p.~414, Nov 2020.

\bibitem{error-mitigation-philip}
P.~Döbler, J.~Pflieger, F.~Jin, H.~D. Raedt, K.~Michielsen, T.~Lippert, and
  M.~S. Jattana, ``Scalable general error mitigation for quantum circuits,''
  2024.

\bibitem{Lieb-Wu-Bethe}
E.~H. Lieb and F.~Y. Wu, ``Absence of mott transition in an exact solution of
  the short-range, one-band model in one dimension,'' {\em Phys. Rev. Lett.},
  vol.~20, pp.~1445--1448, Jun 1968.

\bibitem{jordan-wigner-transform}
M.~A. Nielsen, ``The fermionic canonical commutation relations and the
  jordan–wigner transform.'' Online notes, 2005.
\newblock Available at
  \url{http://futureofmatter.com/assets/fermions_and_jordan_wigner.pdf}.

\bibitem{bravyi-kitaev}
A.~Tranter, S.~Sofia, J.~Seeley, M.~Kaicher, J.~McClean, R.~Babbush, P.~V.
  Coveney, F.~Mintert, F.~Wilhelm, and P.~J. Love, ``The bravyi–kitaev
  transformation: Properties and applications,'' {\em International Journal of
  Quantum Chemistry}, vol.~115, no.~19, pp.~1431--1441, 2015.

\bibitem{bravyi-kitaev2}
J.~T. Seeley, M.~J. Richard, and P.~J. Love, ``The bravyi-kitaev transformation
  for quantum computation of electronic structure,'' {\em The Journal of
  Chemical Physics}, vol.~137, p.~224109, 12 2012.

\bibitem{bethe-ansatz}
H.~Bethe, ``Zur theorie der metalle. i. eigenwerte und eigenfunktionen der
  linearen atomkette,'' {\em Zeitschrift für Physik}, vol.~71, pp.~205--226,
  1931.

\bibitem{karle2021bethe}
A.~Karle, ``Bethe ansatz for the one dimensional heisenberg model,'' project
  report, phys 502 condensed matter physics, Department of Physics and
  Astronomy, University of British Columbia, Dec. 2021.
\newblock Submitted December 14, 2021; available online.

\bibitem{manpreet_heisenberg}
M.~S. Jattana, F.~Jin, H.~De~Raedt, and K.~Michielsen, ``Assessment of the
  variational quantum eigensolver: Application to the heisenberg model,'' {\em
  Frontiers in Physics}, vol.~Volume 10 - 2022, 2022.

\bibitem{layerwise-learning-quantumNN}
A.~Skolik, J.~Mcclean, M.~Mohseni, P.~van~der Smagt, and M.~Leib, ``Layerwise
  learning for quantum neural networks,'' {\em Quantum Machine Intelligence},
  vol.~3, 06 2021.

\bibitem{qiskit2024}
A.~Javadi-Abhari, M.~Treinish, K.~Krsulich, C.~J. Wood, J.~Lishman, J.~Gacon,
  S.~Martiel, P.~D. Nation, L.~S. Bishop, A.~W. Cross, B.~R. Johnson, and J.~M.
  Gambetta, ``Quantum computing with {Q}iskit,'' 2024.

\bibitem{bfgs1}
C.~G. Broyden, ``The convergence of a class of double-rank minimization
  algorithms 1. general considerations,'' {\em IMA Journal of Applied
  Mathematics}, vol.~6, pp.~76--90, 03 1970.

\bibitem{bfgs2}
R.~Fletcher, ``A new approach to variable metric algorithms,'' {\em The
  Computer Journal}, vol.~13, pp.~317--322, 01 1970.

\bibitem{bfgs3}
D.~Goldfarb, ``A family of variable metric methods derived by variational
  means,'' {\em Mathematics of Computation}, vol.~24, pp.~23--26, 1970.

\bibitem{bfgs4}
D.~F. Shanno, ``Conditioning of quasi-newton methods for function
  minimization,'' {\em Mathematics of Computation}, vol.~24, pp.~647--656,
  1970.

\bibitem{SciPy-NMeth}
P.~Virtanen, R.~Gommers, T.~E. Oliphant, M.~Haberland, T.~Reddy, D.~Cournapeau,
  E.~Burovski, P.~Peterson, W.~Weckesser, J.~Bright, S.~J. {van der Walt},
  M.~Brett, J.~Wilson, K.~J. Millman, N.~Mayorov, A.~R.~J. Nelson, E.~Jones,
  R.~Kern, E.~Larson, C.~J. Carey, {\.I}.~Polat, Y.~Feng, E.~W. Moore,
  J.~{VanderPlas}, D.~Laxalde, J.~Perktold, R.~Cimrman, I.~Henriksen, E.~A.
  Quintero, C.~R. Harris, A.~M. Archibald, A.~H. Ribeiro, F.~Pedregosa, P.~{van
  Mulbregt}, and {SciPy 1.0 Contributors}, ``{{SciPy} 1.0: Fundamental
  Algorithms for Scientific Computing in Python},'' {\em Nature Methods},
  vol.~17, pp.~261--272, 2020.

\bibitem{quantum-algos-for-many-body-physics}
Z.~Jiang, K.~J. Sung, K.~Kechedzhi, V.~N. Smelyanskiy, and S.~Boixo, ``Quantum
  algorithms to simulate many-body physics of correlated fermions,'' {\em Phys.
  Rev. Appl.}, vol.~9, p.~044036, Apr 2018.

\bibitem{solving-fermi-hubbard}
C.~Cade, L.~Mineh, A.~Montanaro, and S.~Stanisic, ``Strategies for solving the
  fermi-hubbard model on near-term quantum computers,'' {\em Phys. Rev. B},
  vol.~102, p.~235122, Dec 2020.

\end{thebibliography}

\end{document}